\documentstyle[11pt,newpasp,twoside,epsf]{article}
\begin{document}
\title{The Fate of Intracluster Radio Plasma} \author{Torsten
 A. En{\ss}lin} 
\affil{Max-Planck-Institut f\"{u}r Astrophysik,
 Karl-Schwarzschild-Str.1, Postfach 1317, 85741 Garching, Germany}
%\author{Ima Co-Author}
%\affil{The Name of My Institution, The Full Address of My Institution}

\begin{abstract}
Radio plasma injected by active radio galaxies into clusters of
galaxies quickly becomes invisible due to radiative losses of the
relativistic electrons. In this talk, the fate of radio plasma and its
role for the galaxy cluster is discussed: buoyancy removes it from the
central regions and allows to transfer its energy into the ambient
gas. The remaining low energy electron populations are still able to
emit a low luminosity glow of observable radiation via
synchrotron-self Comptonized emission. Shock waves in the ambient gas
can re-ignite the radio emission.
\end{abstract}

\section{Intracluster radio plasma}

The jets of powerful radio galaxies inflate large cavities in the
intracluster medium (ICM) that are filled with relativistic particles and
magnetic fields. Synchrotron emission at radio frequencies reveals the
presence of electrons with several GeV energies. These electrons have
radiative lifetimes of the order of 100 Myr before their observable
radio emission extinguishes due to radiative energy losses. The
remnants of radio galaxies and quasars are called \lq fossil radio
plasma' or a \lq radio ghosts' ({En{\ss}lin} 1999).
%\cite{1999dtrp.conf..275E}
Their existence as a separate component of the ICM is supported by the
detections of cavities in the X-ray emitting galaxy cluster gas
({B{\"o}hringer} et al. 1993; {Carilli}, {Perley}, \& {Harris} 1994;
{Huang}, \& {Sarazin} 1998; {McNamara} et al. 2000; {Fabian} et
al. 2000; {Finoguenov} \& {Jones} 2001; {Fabian} 2001; {McNamara}
2000; {Heinz} et al. 2001; {Schindler} et al. 2001; and others).
%\cite[and others]{1993MNRAS.264L..25B, 1994MNRAS.270..173C,
%1998ApJ...496..728H, 2000ApJ...534L.135M, 2000MNRAS.318L..65F,
%2001ApJ...547L.107F, fabian2001moriond, McNamara2000Paris,
%heinz2001,schindler2001}
In many cases associated radio emission and in a few cases a lack of
such emission was found, as expected for aging bubbles of radio
plasma. Such bubbles should be very buoyant and therefore rise in the
atmosphere of a galaxy cluster (Churazov et al. 2001). It is not clear
yet if they break into pieces during their ascent and thereby are
slowed down. Another possibility is that they are able to ascend up to
the accretion shock of a galaxy cluster, where their further rise will
be prohibited by the infalling gas of the accretion onto the cluster.

\begin{figure}[t]
\plotone{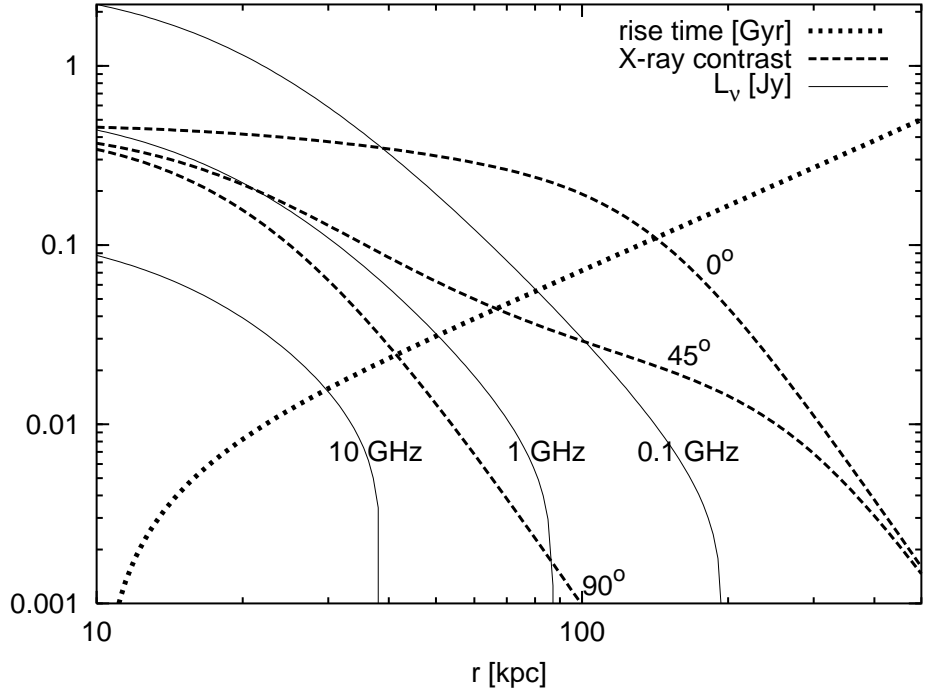}
\caption{Bubble's central X-ray contrast (compared to the undisturbed
cluster) for various angles between plane of sky and Bubble's
trajectory, its radio flux, and its rising time as a function of the
(unprojected) radial position. An X-ray background with 1/100 of the
central cluster surface brightness is assumed, which is responsible
for the strong decrease at large radii of the X-ray contrast in the
$0^\circ$ and $45^\circ$ cases.}
\label{eps5}
\end{figure}

\section{Rising bubbles}

A rough estimate of the rise time of a bubble in a cluster potential
can be obtained by balancing the buoyancy and the hydrodynamical
drag forces (e.g. En{\ss}lin \& Heinz 2002 for further details). For a
relativistic equation of state, which describes the expansion of the
bubble due to the decreasing environmental pressure, such a bubble's
rise is illustrated in Fig. \ref{eps5}. The parameters chosen in this
example are similar to the ones expected for one of the radio lobes of
Perseus A.

Also the synchrotron luminosity at three different frequencies is
shown in Fig. \ref{eps5}. Adiabatic losses dominate the decrease in
radio luminosity, until the synchrotron and inverse Compton losses
have removed all high energy electrons, and the sources fades
completely. This happens first at higher frequencies due to the
higher required electron energies necessary for high frequency
synchrotron emission.

\begin{figure}[t]
\plotone{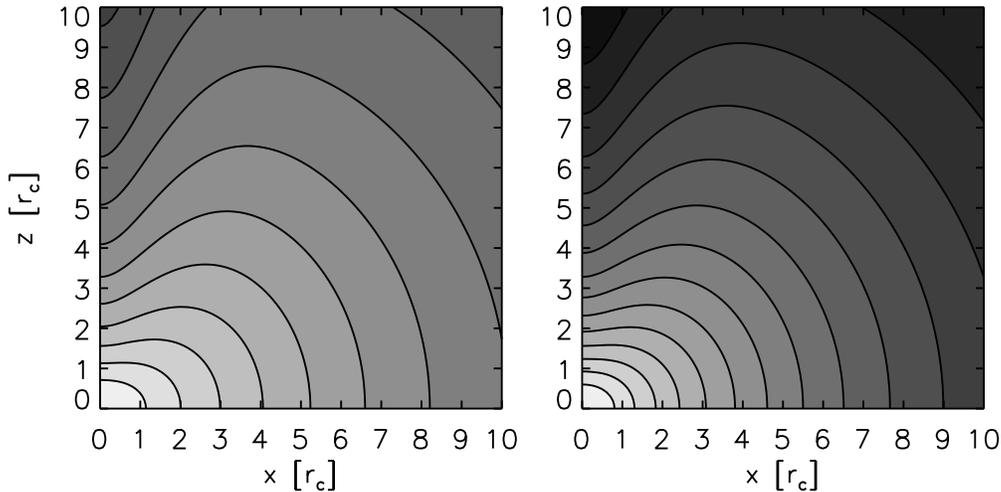}
\caption{Contours of the X-ray deficit significance $({\mathcal
S}/{\mathcal N})$ of a bubble in a galaxy cluster. The contours mark
locations at which a bubble has a significance which is lower by
powers of 2 than its significance if located at the cluster center (in
the lower left corner of each figure, vertical axis is parallel to the
line of sight). Left: the bubble volume expands adiabatically (with a
relativistic equation of state) with the pressure of the isothermal
cluster. Right: the bubble volume is assumed to be independent of
location (incompressible bubble, as a toy model). In both figures, an
X-ray background with 1/100 of the central cluster surface brightness
is assumed. For further details see En{\ss}lin \& Heinz (2002).}
\label{eps6}
\end{figure}

Further, the X-ray contrast of the bubble is shown in Fig. \ref{eps5}
for trajectories along the line of sight ($0^\circ$), within the plane
of sky ($90^\circ$), and $45^\circ$ in between of these two
directions. From this it is clear, that the detectability of the
bubble strongly depends on its 3-dimensional position within the
cluster. 

This is also illustrated in Fig \ref{eps6}, where the signal to noise
ratio of an X-ray cavity at different positions in the cluster is
displayed. The signal is the number of missing photons from the volume
occupied by the cavity, and the noise is given by the typical photon
number fluctuations for an undisturbed region (without cavity) at a
comparable cluster radius and geometry (noise = the square root of the
expected photon number). As can be seen there, the strongest signal to
noise ratio is expected for a cavity within the central plane of the
cluster. This implies, that the yet detected cavities are only a
subsample of the cavities present, and therefore a larger subvolume of
the ICM is likely be permeated by radio plasma than the present day
maps indicate.

\begin{figure}[t]
\plotone{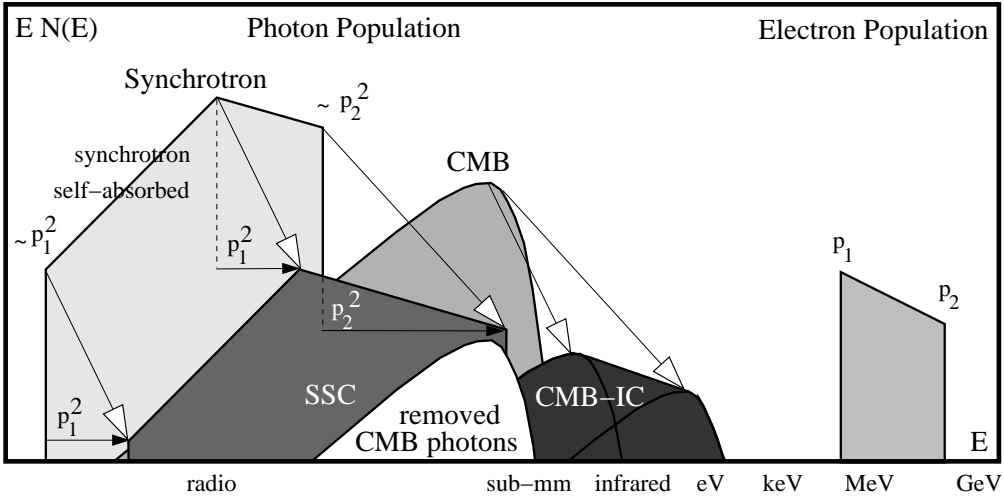}
\caption{Sketch of the SSC and the CMB-IC process.}
\label{eps3}
\end{figure}

\begin{figure}%[t]
\plotone{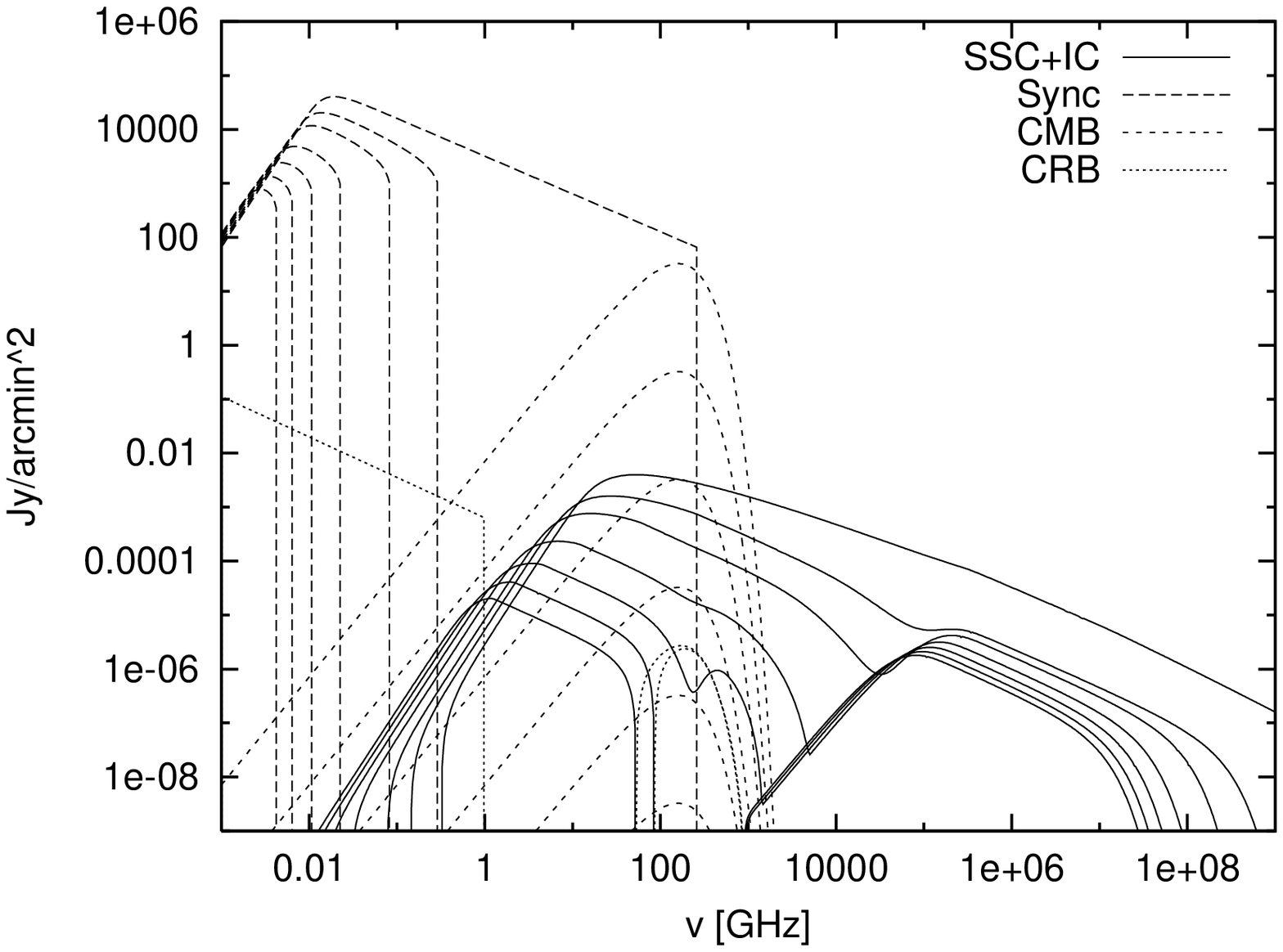}
\caption{Central surface brightness of a Cygnus A-like radio cocoon in
a cooling and expanding phase . The synchrotron (long-dashed) and
SSC+IC spectra (solid) are shown for the stages at the jet-power
shutdown and for later stages (from top to bottom spectra at ages of
0, 20, 40, 80, 120, 160, and 200 Myr are displayed). In spectral
regions, where the SSC+IC processes lead to a reduction of the
brightness below the CMB brightness, the absolute value of the
(negative) SSC+IC surface brightness is plotted by a dotted line.  The
top one of the short-dashed lines is the CMB spectrum, the
short-dashed lines below this are $10^{-2}, 10^{-4}, 10^{-6}, 10^{-8},
\,\mbox{and}\,10^{-10}$ times the CMB spectrum for comparison of the
source to the CMB brightness. The dotted power-law line at frequencies
below 1 GHz is the cosmic radio background (CRB)}
\label{eps4}
\end{figure}

\section{SSC \& CMB-IC}

Even very old radio plasma may be detectable by its long lasting
very low frequency radio emission (kHz -- MHz). Even if this emission
is undetectable directly for terrestrial telescopes, it can be
measured indirectly due to the unavoidable inverse Compton (IC)
scattering of the synchrotron photons by their source electrons
(En{\ss}lin \& Sunyaev 2002). These synchrotron self-Comptonized (SSC)
photons have much higher energies and can therefore be in observable
wavebands.  The relativistic electron population also up-scatters
every other present photon field. Photons of the cosmic microwave
background (CMB) but also of the cosmic radio background (CRB) are
removed from their original spectral location and shifted to much higher
frequencies by IC encounters with the fossils radio plasma electron
population. Since the frequency shift is large for IC scattering by
ultra-relativistic electrons, the CMB flux is decremented within the
whole typical CMB frequency range ({En{\ss}lin} \& {Kaiser} 2001).
%\cite{ensslin2000a}.  
The spectral signature of all these processes are sketched in
Fig. \ref{eps3}.

The strongest detectable sources of such low frequency SSC emission should be
radio lobes of just extinguished powerful radio galaxies (see
Fig. \ref{eps4} and En{\ss}lin \& Sunyaev (2002)
%\cite{EnsslinSunyaevI01} 
for details). If for example the central engine of Cygnus A would
decease today, its directly observable radio lobe synchrotron emission
would vanish within some 10 Myr due to radiation and adiabatic losses
of the (still) expanding radio plasma. But SSC and CMB-IC emission (or
decrement) can remain for a few 100 Myr. Since the SSC emission is
very sensitive to the compression state of the radio plasma it would
decrease rapidly due to adiabatic expansion during the buoyant rise of
the radio bubble in the cluster atmosphere. The CMB-IC process is much
less sensitive to compression and would start to dominate the spectrum
above 30 GHz after roughly 100 Myr.

The detection of SSC from radio ghosts is an observational
challenge. It would be rewarded by revealing the locations of fossil
radio plasma graveyards. It would provide important information on the
lower end of the relativistic electron population. This would be very
valuable since the electron energy range above a few 10 keV and below
100 MeV is still an unexploited spectral regions. Further, due to the
strong dependence of the SSC emission on the compression stage of
radio plasma, SSC emission is also a sensitive probe of the ICM
pressure.

Due to its broad frequency spectra, it can be probed with several
future high sensitivity instruments, ranging from lowest radio
frequency radio telescopes as GMRT and LOFAR, over microwave
spacecrafts like MAP and PLANCK, balloon and ground based CMB
experiments, and sub-mm/IR projects as ALMA and the HERSCHEL
satellites. A multi-frequency sky survey, as will be provided by the
Planck experiment, should allow to search for the SSC and relativistic
IC spectral signature of many nearby clusters of galaxies and radio
galaxies, at least in a statistical sense by co-adding the signals
from similar sources.  In addition to this, there should be targeted
observations of promising candidates, as e.g. the recently reported
X-ray cluster cavities without apparent observable synchrotron
emission. New and upcoming radio telescopes like LOFAR, GMRT, EVLA,
ATA, and SKA should have a fairly good chance to detect such
sources. E.g. a Cygnus-A like radio cocoon should be detectable for
these telescopes out to a few 100 Mpc even $\sim$ 100 Myr after the
jetpower shutdown (En{\ss}lin \& Sunyaev 2002).
%\cite{EnsslinSunyaevI01}.

\begin{figure}[t]
\plotone{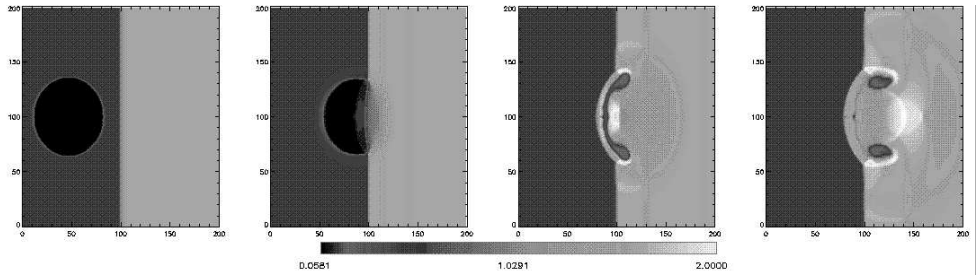}
\caption{Shock passage of a hot, magnetized bubble (a radio ghost)
through a shock wave. The flow goes from the left to the right. The
evolution of the mid plane gas density is displayed (white is dense,
black is dilute gas)}
\label{eps1}
\end{figure}

\section{Shock wave re-illumination}

Whenever a radio ghosts is hit by a shock wave, which may originate
either from a cluster merger or from the steady accretion of gas onto
the still forming large scale structure, it is strongly
compressed. The compression should be adiabatic since typical ICM
shock speeds of a few 1000 km/s are expected to be well below the
internal sound speed of the fossil but still relativistic
plasma. Since the radio plasma has to adapt to the new ambient
pressure, the compression factor can be high and the particles and
magnetic fields can gain a substantial amount of energy. The
synchrotron emission can go up by a large factor, especially at
frequencies which were only a little bit higher than the cutoff
frequency of the uncompressed fossil plasma
({En{\ss}lin} \& {Gopal-Krishna} 2001).
%\cite{2001A&A...366...26E}. 
Thus, the radio plasma can be revived to emit at observing frequencies
if it was not too old, a few 100 Myr inside and a few Gyr at the
boundary of galaxy clusters.

\begin{figure}[t]
\plotone{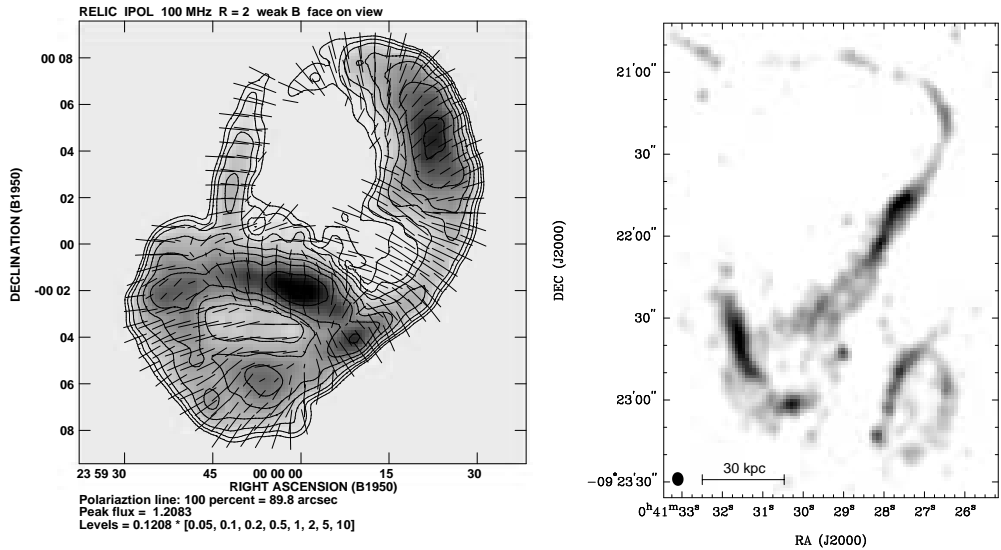}
\caption{Left: synthetic radio map of a shocked radio ghost
({En{\ss}lin} \& {Br{\"u}ggen} 2002).
%\cite{ensslinbrueggen01}. 
Right:
observed cluster radio relic in Abell 85 at 1.4 GHz (Slee et
al. 2001)
%\cite{slee2001}
}
\label{eps2}
\end{figure}

3-D magneto-hydrodynamical simulations ({En{\ss}lin} \&
{Br{\"u}ggen} 2002)
%\cite{ensslinbrueggen01} 
show that during the traversal of a shock wave, the radio plasma is
first flattened and then breaks up into small filaments, often in form
of one or several tori. The formation of a torus can be seen in
Fig. \ref{eps1}. At places where the hot, under-dense radio plasma
bubble touches the shock wave the balance of pre-shock ram-pressure
and post-shock thermal pressure disappears due to the lack of
substantial mass load of the advected radio plasma. The post shock gas
therefore starts to break through the radio plasma and finally
disrupts it into a torus or a more complicated filamentary structure.

The magnetic fields becomes mostly aligned with the filaments
leading to a characteristic polarization signature which can be seen in
the synthetic radio map displayed in Fig. \ref{eps2}.

Polarized radio emitting regions of often filamentary morphologies
could be found in a (recently strongly growing) number of merging
clusters of galaxies. In most cases they are near those places where
shock waves are expected from either observed temperature structures
or comparison of X-ray maps to simulated cluster merger. These radio
sources are called cluster radio relics 
({Feretti} 1999, and references therein).
%\cite[and references therein]{1999dtrp.conf....3F}.  
An observed radio map of a filamentary
cluster radio relic in Abell 85 is also displayed in Fig. \ref{eps2}
for comparison. Lower frequency observation show that the upper
filament of this relic forms (at least in projection) a closed torus
({Giovannini} \& L.~{Feretti} 2000).
%\cite{2000NewA....5..335G}.

Thus, sensitive observation of cluster radio relics are able to probe
several properties of ICM shock waves.  Since the major diameter of
the fossil radio plasma is approximately conserved, a rough estimate
of the compression factor can be derived from relic radio maps by
measuring the filament diameters. The compression factor depends only
on the pressure jump in the shock and the equation of state of the
radio plasma. Therefore, the shock strength is measurable for a given
radio plasma equation of state. Or, if detailed X-ray maps allow to
estimate the shock strength independently, the equation of state of
radio plasma can be measured.  Furthermore, the total radio
polarization of cluster radio relics (after averaging over the source)
contains in principle enough information to entangle the 3-D
orientation of the shock wave: The sky-projected electric vector is
aligned with the projected shock normal. The polarization fraction is
highly correlated with the angle between the shock normal and the line
of sight ({En{\ss}lin} \& {Br{\"u}ggen} 2002).
%\cite{ensslinbrueggen01}.

\end{document}